\documentclass[11pt]{article}
\usepackage{moriond,epsfig}
\usepackage{amssymb,amsthm,amsfonts}

\def\Journal#1#2#3#4{{#1} {\bf #2}, #3 (#4)}

\def\NPB{{\em Nucl. Phys.} B}
\def\PLB{{\em Phys. Lett.}  B}
\def\PRL{\em Phys. Rev. Lett.}
\def\PRD{{\em Phys. Rev.} D}

\def\JHEP{\em JHEP}
\def\MPLA{{\em Mod. Phys. Lett.} A}

\def\beq{\begin{equation}}
\def\eeq{\end{equation}}
\def\bea{\begin{eqnarray}}
\def\eea{\end{eqnarray}}

\begin{document}

\begin{flushright}
IFT-05-17\\
SACLAY-T05/117
\end{flushright}

\vspace*{2cm}

\title{FLAVOUR CHANGING NEUTRAL CURRENTS AND INVERTED
SFERMION MASS HIERARCHY
\footnote{Talk given by S.L. at the XLth Rencontres de Moriond on Electroweak
Interactions and Unified Theories, La Thuile, Aosta Valley, Italy, 5-12 March 2005.}}

\author{ P.H. CHANKOWSKI,$\, ^*$ K. KOWALSKA,$\, ^*$ S. LAVIGNAC$\, ^\#$
and S. POKORSKI$^{\, *,\dagger}$  }

\address{$^*$ Institute of Theoretical Physics, Warsaw University, Ho\.za 69, 00-681, 
Warsaw, Poland\\
$^\#$ Service de Physique Th\'eorique, CEA-Saclay, F-91191 Gif-sur-Yvette
Cedex, France \footnote{Unit\'e de Recherche associ\'ee au CNRS (URA 2306).}\\
$^\dagger$ Theory Division, Physics Department, CERN, CH-1211 
Geneva 23, Switzerland}

\maketitle\abstracts{
We study the contraints on non-flavour-blind soft supersymmetry
breaking terms coming from flavour and CP violating processes
in the presence of hierarchical Yukawa couplings, and quantify 
how much these constraints are weakened in the regions of the MSSM
parameter space characterized by heavy gauginos and multi-TeV
sfermion masses, respectively. We also study the inverted sfermion
mass hierarchy scenario
in the context of $D$-term supersymmetry breaking,
and show that generic hierarchical Yukawa couplings with
arbitrary phases require first generation squarks in the few $10$ TeV
range.
}

\section{Introduction}

In contrast with the successful predictions of the Standard Model,
supersymmetry does not guarantee suppressed flavour changing
neutral current (FCNC) and CP violating processes. Instead the
predictions of supersymmetric models depend on two sets of
unknown parameters, the soft supersymmetry breaking terms
and the Yukawa coupling matrices.
Arbitrary phases and flavour structures in these parameters
would lead to unacceptably large contributions to FCNC and
CP violating observables: this is the well-known supersymmetric
flavour problem.
At the phenomenological level, there are well-known radical
possibilities that make the above mentioned dependence trivial,
and the Standard Model predictions are recovered:
flavour universal soft terms~\cite{ELNA},
or alignment of the sfermion and fermion mass matrices~\cite{NISE}.
However, the mechanism of spontaneous supersymmetry
breaking and its transmission to the visible sector is unkown and ultimately
may not be consistent with any of these options (measurements of the
supersymmetric mass spectrum at the LHC may tell us).

It is therefore useful to study the dependence of the FCNC
and CP constraints on the supersymmetric mass spectrum
and on the pattern of Yukawa couplings.
In this work, we perform such an analysis in the framework of supergravity-mediated 
supersymmetry breaking and in the presence of hierarchical Yukawa couplings,
without insisting on keeping a moderate fine-tuning in the Higgs potential.
In particular, we quantify and compare the FCNC and CP problems in two regions
of the supersymmetric parameter space where they are expected to be less
stringent: the region with sub-TeV (GUT-scale) sfermion masses but
possibly heavy gauginos, on the one hand, and the region with sfermion masses
in the multi-TeV range, on the other hand.
We also study the so-called ``inverted hierarchy'' scenario~\cite{DUPOSA,COKANE},
with heavy first two generation sfermions and lighter third generation sfermions,
which has been put forward as a way to suppress the most dangerous flavour
and CP violating processes while keeping the fine-tuning at an acceptable level.
We investigate the predictions of this scenario for FCNC processes
in a specific inverted hierarchy model with $D$-term supersymmetry breaking.


\section{FCNC and CP constraints: heavy gauginos versus heavy sfermions}
\label{sec:FCNC}

The goal of this section is to quantitify and compare the FCNC and CP problems
in the two regions of the supersymmetric parameter space where the (GUT-scale)
sfermion masses are in the sub-TeV range but the gauginos are relatively heavy,
and where the sfermion masses are in the multi-TeV range, respectively.

Let us first specify our assumptions about soft sfermion masses and Yukawa
couplings. In the Yukawa sector, we consider matrices of the hierarchical type,
for which the smallness of the charged fermion mass ratios
result from a hierarchy among Yukawa couplings rather than from cancellations
between large entries in the Yukawa matrices.
For definiteness, we shall use in our subsequent numerical study one of the quark
and lepton Yukawa textures that were analyzed in Ref.~\cite{CHKOLAPO}.
These textures are associated with the spontaneous breaking, close to the
$GUT$ scale, of a horizontal $U(1)$ symmetry~\cite{FRONI}
and have the following structure:
$\mathbf{Y}_u^{AB} = C_u^{AB} ~\epsilon^{q_A + \bar u_B + h_u}$,
$\mathbf{Y}_d^{AB} = C_d^{AB} ~\epsilon^{q_A + \bar d_B + h_d}$ and
$\mathbf{Y}_e^{AB} = C_e^{AB} ~\epsilon^{l_A + \bar e_B + h_d}$,
 where the $C_{u,d,e}^{AB}$ are arbitrary complex coefficients of order one,
$\epsilon \ll 1$ is a small parameter associated with the spontaneous breaking
of the horizontal symmetry, and
$q_A$, $\bar u_A$, $\bar d_A$, $l_A$, $\bar e_A$, $h_u$, $h_d$ stand
for the horizontal charges (assumed to be positive) of the MSSM superfields
$Q_A$, $U^c_A$, $D^c_A$, $L_A$, $E^c_A$, $H_u$, $H_d$.
Specifically, we choose model 1 from Ref.~\cite{CHKOLAPO}, with horizontal charges
$q_A = \bar u_A = \bar e_A = (3,2,0)$, $l_A = \bar d_A = (4,2,2)$ and $h_u = h_d = 0$.
Since the symmetry breaking parameter $\epsilon$ turns out to be very close
numerically to the Cabbibo angle $\lambda \simeq 0.22$, we set $\epsilon = \lambda$.
The associated quark and charged lepton Yukawa matrices then read:
\beq
  \mathbf{Y}_u \sim y_t \left(
    \begin{array}{ccc} \lambda^6 & \lambda^5 & \lambda^3  \\
                                     \lambda^5 & \lambda^4 & \lambda^2  \\
                                     \lambda^3 & \lambda^2 & 1
    \end{array} \right) ,  \quad
  \mathbf{Y}_d \sim y_b \left(
    \begin{array}{ccc} \lambda^5 & \lambda^3 & \lambda^3  \\
                                     \lambda^4 & \lambda^2 & \lambda^2  \\
                                     \lambda^2 & 1 & 1
    \end{array} \right) ,  \quad
  \mathbf{Y}_e \sim y_\tau \left(
    \begin{array}{ccc} \lambda^5 & \lambda^4 & \lambda^2  \\
                                     \lambda^3 & \lambda^2 & 1  \\
                                     \lambda^3 & \lambda^2 & 1
    \end{array} \right) ,
\label{eq:Y_ude}
\eeq
where the symbol $\sim$ reminds us that an order one complex factor is understood
in each entry of the matrices $\mathbf{Y}_{u,d,e}$. In practice, we shall use randomly
generated sets of order one coefficients \{$C_u^{AB}$, $C_d^{AB}$, $C_e^{AB}$\}
that fit the measured values of the quark and lepton masses and mixings, with
renormalization group evolution of the Yukawa couplings between the GUT scale
and the weak scale taken into account~\cite{CHKOLAPO}.
In the neutrino sector, this charge assignment also yields correct agreement
with oscillation data upon adjusting the relevant order one
parameters~\cite{CHKOLAPO,ALFEMA}.

In the supersymmetry breaking sector, we do not rely on any specific model
and make only mild assumptions about the flavour structure
of the soft terms. In particular, we consider soft sfermion masses
with both splittings among entries on the diagonal and
non-vanishing off-diagonal entries (suppressed by some power of $\lambda$):
\beq
  m^2_Q, m^2_U, m^2_D, m^2_L, m^2_E\ =\
      \left(\matrix{m^2_1 & {\cal O} (m^2 \lambda^n) & {\cal O} (m^2 \lambda^n) \cr
                            {\cal O} (m^2 \lambda^n) & m^2_2 & {\cal O} (m^2 \lambda^n) \cr
                            {\cal O} (m^2 \lambda^n) & {\cal O} (m^2 \lambda^n) & m^2_3}\right) ,
\label{eq:M2_QudLe}
\eeq
where, unless otherwise stated, $m^2_1, m^2_2, m^2_3 \sim m^2$, and we have
assumed the same level of suppression for all off-diagonal entries.
For the $A$-terms, we assume the following flavour structure:
\beq
  A^{AB}_u\, =\, a^{AB}_u \lambda^{q_A + \bar u_B + h_u} A_0\ ,  \quad
  A^{AB}_d\, =\, a^{AB}_d \lambda^{q_A + \bar d_B + h_d} A_0\ ,  \quad
  A^{AB}_e\, =\, a^{AB}_e \lambda^{l_A + \bar e_B + h_d} A_0\ ,
\label{eq:A_ude}
\eeq
where $A_0$ is a mass scale and the $a^{AB}_{u,d,e}$ are complex order one coefficients.
While compatible with the horizontal symmetry manifest in the fermion sector,
this ansatz departs from the standard proportionality assumption $A^{AB} = A_0 Y^{AB}$,
which would correspond to $a^{AB}_{u,d,e} = C^{AB}_{u,d,e}$.

To estimate the contributions to flavour and CP violating processes due to the
hierarchical mass matrices (\ref{eq:Y_ude}) -- (\ref{eq:A_ude}), we use the standard
mass insertion parameters~\cite{HAKORA} (here in matrix form):
\beq
  \delta^d_{LL}\ \equiv\ \frac{D_L m^2_Q D^\dagger_L}{\bar m^2_{\tilde d}}\ , \quad
  \delta^d_{RR}\ \equiv\ \frac{D_R m^2_D D^\dagger_R}{\bar m^2_{\tilde d}}\ , \quad
  \delta^d_{LR}\ \equiv\ \frac{D_L \tilde m^{d 2}_{LR} D^\dagger_R}{\bar m^2_{\tilde d}}\ ,
\eeq
where $D_L$ (resp. $D_R$) is the rotation that brings the left-handed (resp.
right-handed) down quarks to their mass eigenstate basis,
$\tilde m^{d 2}_{LR} =  (A_d - \mu \tan \beta\, Y_d)\, v_d$,
and $\bar m_{\tilde d}$ is an average down squark mass.
The diagonal terms, irrelevant to flavour violation, have been omitted
in $\delta^d_{LL}$ and $\delta^d_{RR}$. Off-diagonal entries of order $\lambda^n$
in the squark soft mass matrices will give contributions of order $\lambda^n$
to the corresponding \footnote{Needless to say,
a non-vanishing ($1,3$) entry in $m^2_Q$, for instance, also gives a contribution
to $(\delta^d_{LL})^{12}$, but with an additional suppression by the rotation angle
$D^{23}_L$.} $(\delta^d_{LL})^{AB}$'s and $(\delta^d_{RR})^{AB}$'s,
while the contribution of splittings among diagonal entries strongly depends
on the magnitude of the mixing in the Yukawa textures. For example,
$m^2_1 \neq m^2_2$ in $m^2_Q$ gives a
contribution $D^{12}_L D^{22 \star}_L (m^2_2 - m^2_1) / \bar m^2_{\tilde d}
\sim \lambda\, (m^2_2 - m^2_1) / \bar m^2_{\tilde d}$ to
$(\delta^d_{LL})^{12}$. Finally, Eq.~(\ref{eq:A_ude}) gives
$(\delta^d_{LR})^{AB} \sim (A_0 v_d / m^2) \lambda^{q_A+\bar d_B+h_d}$
($A \neq B$)
and \footnote{We assume here that the phases of the common gaugino mass
and of the $\mu$ parameter can be simultaneously rotated away.}
$|\mbox{Im} (\delta^d_{LR})^{AA}| \lesssim  |A_0|\, m_{d_A} / m^2$.
Analogous quantities $(\delta^u_{MN})^{AB}$ and $(\delta^e_{MN})^{AB}$, where
$M, N = L$ or $R$ are defined in the up squark and slepton sectors, respectively.

Let us discuss in greater detail the expected magnitude of the FCNC processes
induced by splittings among diagonal entries of the sfermion soft mass matrices.
The diagonalization of the Yukawa matrices~(\ref{eq:Y_ude}) yields the following
hierarchical structures for the rotations that bring the quarks and charged leptons
to their mass eigenstate basis:
\bea
  \mathbf{U}_L\, ,\, \mathbf{D}_L\, ,\, \mathbf{U}_R\, ,\, \mathbf{E}_R \sim \left(
    \begin{array}{ccc} 1 & \lambda & \lambda^3  \\
                                     \lambda & 1 & \lambda^2  \\
                                     \lambda^3 & \lambda^2 & 1
    \end{array} \right) ,  \quad
  \mathbf{D}_R\, , \, \mathbf{E}_L \sim \left(
    \begin{array}{ccc} 1 & \lambda^2 & \lambda^2  \\
                                     \lambda^2 & 1 & 1  \\
                                     \lambda^2 & 1 & 1
    \end{array} \right) .
\label{eq:R_ude}
\eea
Let us first consider the implications of Eq.~(\ref{eq:R_ude}) in the down squark sector.
Since $D_L$ has the same hierarchical structure as the CKM matrix
$V = U^\dagger_L D_L$,
the order of magnitude of the dominant contribution to $(\delta^d_{LL})^{AB}$ is:
\begin{equation}
  (\delta^d_{LL})^{AB}\ \sim\ V^{AB} (m^2_B - m^2_A) / m^2\ ,
\end{equation}
Hence $m^2_A \neq m^2_B$ induces a large
$(\delta^d_{LL})^{12}$, but a smaller $(\delta^d_{LL})^{13}$ or
$(\delta^d_{LL})^{23}$. Furthermore, $(\delta^d_{RR})^{12}$ and
$(\delta^d_{RR})^{13}$ are of order $\lambda^2 (m^2_B - m^2_A) / m^2$, while
$(\delta^d_{RR})^{23} \sim (m^2_3 - m^2_2) / m^2$.
We therefore expect large contributions from splittings among diagonal entries
in the squark mass matrices to $\Delta m_K$ and to $\epsilon_K$ (since large
phases are generally present in the rotation
matrices $D_L$ and $D_R$), while the contributions to $B^0_d - \bar B^0_d$ mixing
should be suppressed. Large effects are also expected in $b \rightarrow s$ transitions,
due to $(\delta^d_{RR})^{23}$,
as well as in $D^0$-$\bar D^0$ mixing (including CP violating
effects, which are absent in the Standard Model) and in $\mu \rightarrow e \gamma$,
since Eq.~(\ref{eq:R_ude}) implies $U^{12}_L \sim U^{12}_R \sim E^{12}_R
\sim D^{12}_L \sim \lambda$, as well as $E^{12}_L \sim D^{12}_R \sim \lambda^2$.
Finally, since $E^{23}_R \sim D^{23}_R\sim 1$, $\tau \rightarrow \mu \gamma$
may also receive sizeable contributions from splittings among the diagonal
entries of $m^2_E$.

So far our discussion has been qualitative and has ignored the effect of the
renormalization group evolution on soft supersymmetry breaking terms, as well
as the dependence of the flavour and CP violation observables on the superparticle
mass spectrum. We now move to the quantitative study of a few observables
in the low ($\lesssim 1$ TeV)  and the high ($5-50$ TeV) sfermion mass regions for
various values of the GUT-scale parameters $A_0$ and $M_{1/2}$, $|\mu|$ and
$B \mu$ being determined from radiative electroweak symmetry breaking. Our goal is
to compare the FCNC and CP problems in these two regions for the
hierarchical Yukawa couplings~(\ref{eq:Y_ude}). In order to do this, we scan
over the different flavour and CP violating (GUT-scale) parameters in the sfermion
sector \footnote{We do not include in this list the right-handed neutrino couplings
needed for the seesaw mechanism, which are known to induce large
flavour violations in the slepton sector through radiative corrections~\cite{BOMA}.
In fact our charge assignment suppresses the corresponding contribution to the
$l_j \rightarrow l_i \gamma$ amplitude by a factor $\lambda^{l_i+l_j+2 \bar n_3}$,
where $\bar n_3$ is the horizontal charge of the third generation right-handed
neutrino; for $\mu \rightarrow e \gamma$, this is already $\lambda^6 \sim 10^{-4}$
for $\bar n_3=0$, which can safely be neglected.}: (i) the splittings $m^2_A - m^2_B$
among diagonal entries of the soft sfermion mass matrices;
(ii) the off-diagonal entries of the soft sfermion mass matrices (in practice
the complex order one coefficients multiplying some fixed power of $\lambda$);
(iii) the complex order one coefficients $a^{AB}_{u,d,e}$ in the $A$-terms,
Eq.~(\ref{eq:A_ude}). Order one parameters are scanned in the range $[0.3-3]$.
We also vary the Yukawa couplings by using 100 randomly generated sets
of the order one coefficients $C^{AB}_{u,d,e}$ fitting the measured values of the
quark and lepton masses and mixing angles.

\begin{figure}
\begin{center}
  \psfig{figure=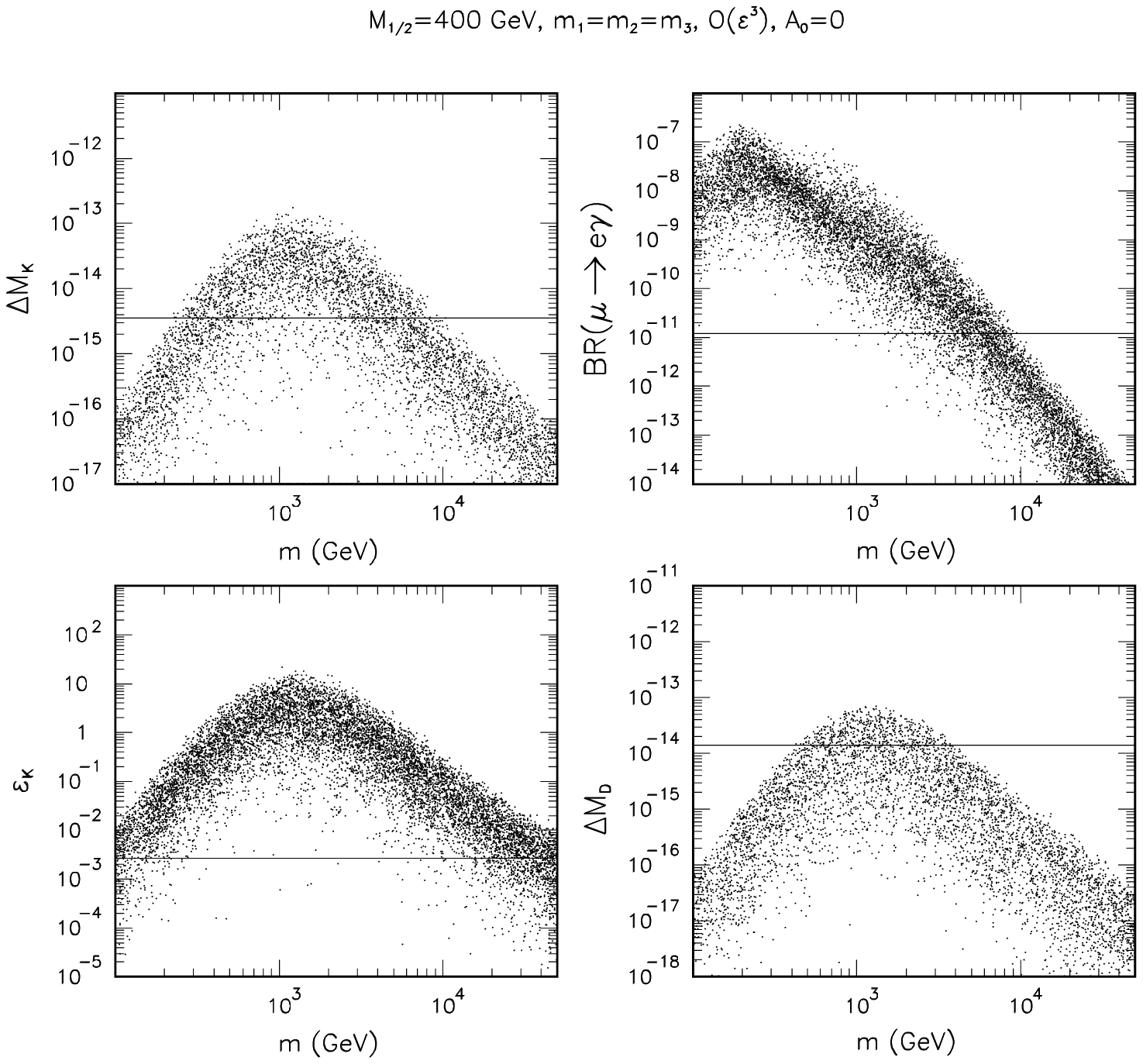,width=0.48\textwidth}
  \psfig{figure=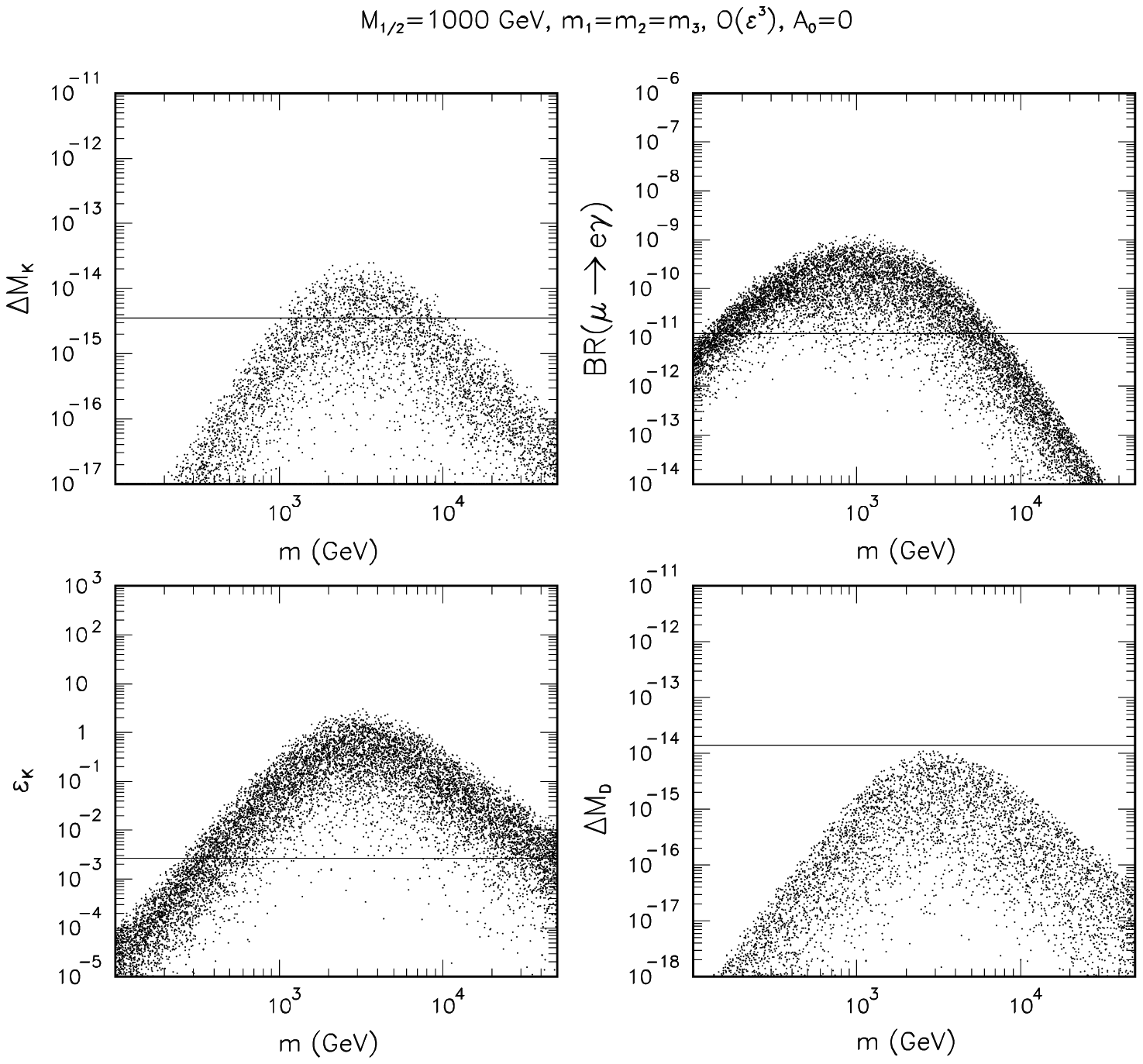,width=0.48\textwidth}
\end{center}
\vskip .1cm
\begin{center}
  (a) \hskip 0.45\textwidth (b)
\end{center}
\caption{$\Delta m_K$ ({\it top left}), $\mbox{BR} (\mu \rightarrow e \gamma)$ ({\it top right}),
$\epsilon_K$ ({\it bottom left}) and $\Delta m_D$ ({\it bottom right}), as a function
of the GUT-scale sfermion mass parameter $m$, assuming $m_1= m_2 = m_3 \equiv m$,
off-diagonal entries of order $\lambda^3$ in the sfermion mass matrices, $A_0 = 0$
and (a) $M_{1/2} = 400$ GeV, (b) $M_{1/2} = 1000$ GeV.
Order one coefficients in the Yukawa matrices and in the off-diagonal entries
of the sfermion mass matrices are varied in the range $[0.3,3]$. \hfill
\label{fig:a0e3g4_10}}
\end{figure}

Let us first consider off-diagonal entries of order $\lambda^3$.
Fig.~\ref{fig:a0e3g4_10} shows $\Delta m_K$, $\epsilon_K$,
$\mbox{BR} (\mu \rightarrow e \gamma)$ and $\Delta m_D$
as a function of the GUT-scale sfermion mass parameter $m$, assuming
$m_1 = m_2 = m_3 \equiv m$, off-diagonal entries of order $\lambda^3$
in the sfermion mass matrices, $A_0 = 0$ and $M_{1/2} = 400$ GeV
and $1000$ GeV, respectively. 
Not surprisingly, the most severe constraints come from $\epsilon_K$ and
$\mu \rightarrow e \gamma$.
For $M_{1/2} = 400$ GeV, the scatter plots for $\Delta m_K$, $\epsilon_K$
and $\Delta m_D$ look approximately symmetric around $m = 1$ TeV,
where these observables reach a maximum. Above $1$ TeV, the effect of
decoupling the squarks running in the loop is clearly visible: $\Delta m_K$,
$\epsilon_K$ and $\Delta m_D$ decrease as $1/ m^2$, as expected. This effect
is practically insensitive to the value of $M_{1/2}$. Below
$1$ TeV, on the other hand, one can see the ``aligning effect'' of the gluino
mass on the squark mass matrices~\cite{BRIBMU} (for a recent discussion
of this effect, see Ref.~\cite{CHLEPO}).
Indeed, the one-loop renormalization group equations of squark masses
receive a large and negative contribution from the gauginos, whose effect
is to enhance the diagonal entries of the squark mass matrices by a universal
piece. Schematically, one has:
\beq
  16 \pi^2\, \frac{d m^2_{Q,U,D}}{d \ln \mu}\ =\ (\mbox{Yukawas})\,
    -\, \sum_i c_i\, g^2_i M^2_i\, \mathbb{I}_{3 \times 3}\, +\, \cdots\ .
\eeq
For the first two generations, the Yukawa couplings can be neglected, implying
$m^2_A (M_Z) \approx m^2_{A} (M_{GUT}) + c\, M^2_{1/2}$ ($A=1,2$), with
$c \approx (6-7)$. Hence, $(\delta^d_{LL})^{12}$ is suppressed with respect
to its ``GUT-scale value'' $\lambda^3$:
$(\delta^d_{LL})^{12} \approx m^2 \lambda^3 / (m^2 + c\, M^2_{1/2})$, and this effect
is more important for larger values of the ratio $M_{1/2} / m$. This is illustrated
in the two sets of plots of Fig.~\ref{fig:a0e3g4_10}, which only differ by the value of
$M_{1/2}$: the supersymmetric contributions to $\Delta m_K$ and $\epsilon_K$
are under much better control, in the low $m$ region, for $M_{1/2} = 1000$ GeV
than for $M_{1/2} = 400$ GeV; in both case, $\Delta m_K$, $\epsilon_K$ and
$\Delta m_D$
decrease quickly when $m$ drops below $1$ TeV. Note that $m = 200$ GeV and
$M_{1/2} = 400$ ($1000$) TeV correspond to $m_{\tilde q_{1,2}} \approx 1$ ($2.5$) TeV
and $M_{\tilde g} \approx 1.2$ ($3$) TeV; thus a GUT-scale sfermion mass parameter
$m$ in the few $100$ GeV range does not necessarily mean light squarks.
In the lepton sector, the aligning effect is much milder,
since $c \approx 0.5$ ($0.15$) for $m^2_L$ ($m^2_E$). It follows that, for moderate
values of $M_{1/2}$, the $\mu \to e \gamma$ constraint becomes more and more
severe when $m$ decreases; for large values of $M_{1/2}$, however,
$\mbox{BR} (\mu \to e \gamma)$ start decreasing when $m$ drops below $1$ TeV.

\begin{figure}
\begin{center}
  \psfig{figure=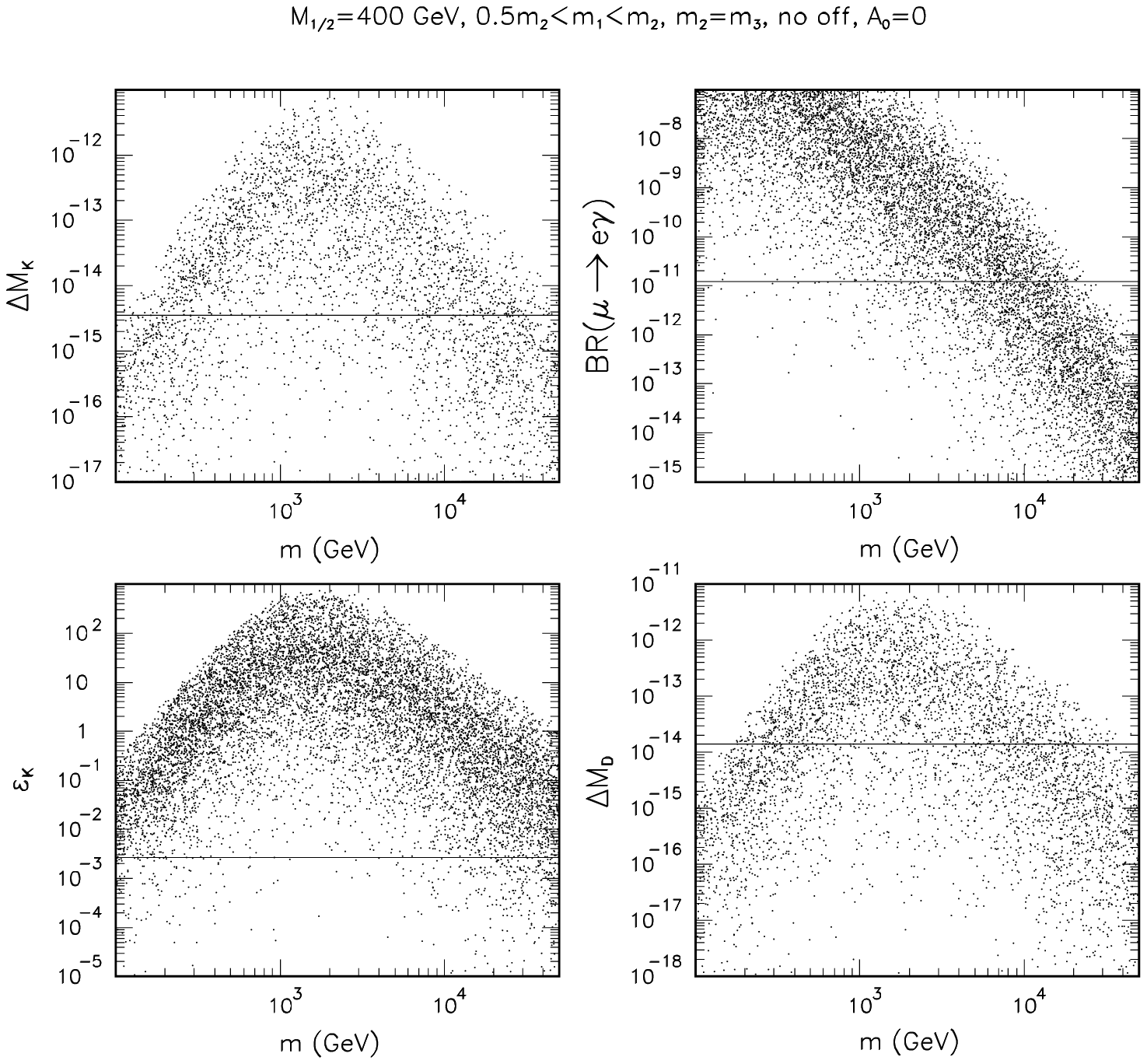,width=0.48\textwidth}
  \psfig{figure=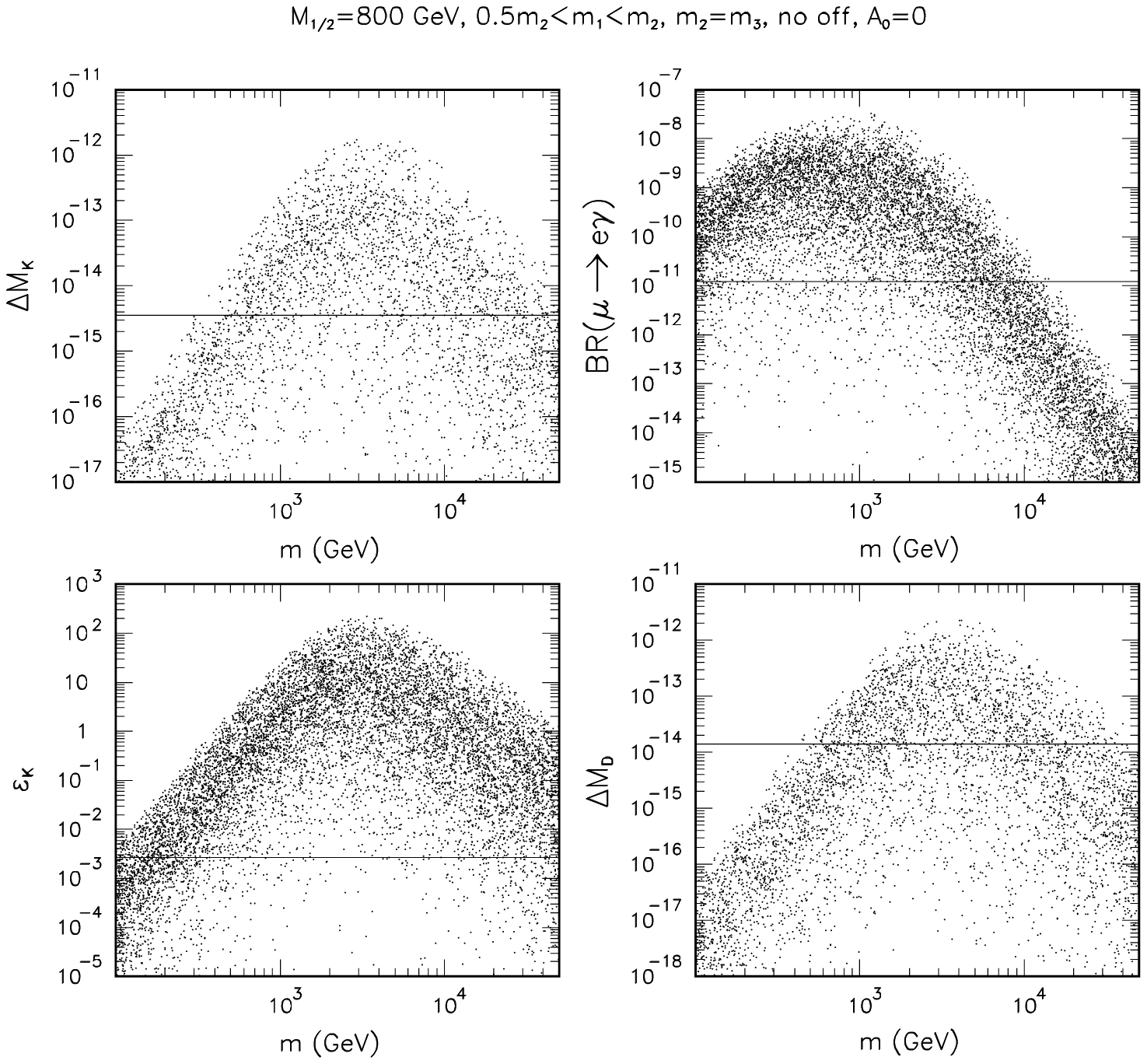,width=0.48\textwidth}
\end{center}
\vskip .1cm
\begin{center}
  (a) \hskip 0.45\textwidth (b)
\end{center}
\caption{Same as Fig.~\ref{fig:a0e3g4_10}, assuming $m_2 = m_3 \equiv m$
and scanning over $m_1$ in the range $[0.5\, m, m]$, with vanishing
off-diagonal entries in the sfermion mass matrices, $A_0 = 0$ and
(a) $M_{1/2} = 400$ GeV, (b) $M_{1/2} = 800$ GeV. \hfill
\label{fig:m1a0r1g4_8}}
\end{figure}

Let us now consider a splitting between the first two diagonal entries
of the sfermion mass matrices. Fig.~\ref{fig:m1a0r1g4_8} shows $\Delta m_K$,
$\epsilon_K$, $\mbox{BR} (\mu \rightarrow e \gamma)$ and $\Delta m_D$
as a function of the GUT-scale sfermion mass parameter $m$, assuming
$m_2 = m_3 \equiv m$ and scanning over $m_1$ in the range $[0.5 m, m]$,
with vanishing off-diagonal entries in the sfermion mass matrices,
$A_0 = 0$ and $M_{1/2} = 400$ GeV and $800$ GeV, respectively.
One can observe qualitatively similar effects to the case of
non-vanishing off-diagonal entries in the low and high $m$ regions.
However, as already mentioned above, there is a substantial difference
between the two sources of flavour violations. While the contribution of
off-diagonal entries to FCNC processes only weakly depends on the
Yukawa couplings, especially if the latter are of the hierarchical type,
this is no so for the contribution of splittings between the diagonal entries.
For the ``generic'' hierarchical case considered in this paper,
$U_L$ and $D_L$ have the same hierarchical structure as the CKM matrix.
In particular, $D^{12}_L$ is of order the Cabibbo angle and gives a large
$(\delta^d_{LL})^{12}$; furthermore, $D^{12}_R$ is only suppressed
by an additional factor of $\lambda$ relative to $D^{12}_L$ (remember that the
strongest constraints from the kaon sector are on the
combination~\cite{GAGAMASI} $(\delta^d_{LL})^{12} (\delta^d_{RR})^{12}$).
Yukawa textures with suppressed mixing
among down-type quarks (either in the left-handed or in the right-handed sector, or
in both) would yield a much smaller contribution to $\Delta m_K$ and $\epsilon_K$,
thus softening the constraint on the $m^2_1 - m^2_2$ splitting in the squark
sector. Similarly, a much stronger hierarchical structure for $Y_e$,
with the large lepton mixing angles coming solely from the neutrino sector,
would at least partially relax the strong constraints from lepton flavour violating
processes such as $\mu \to e \gamma$.

Finally, non-zero $A$-terms have little impact on FCNCs, due to their
proportionality to Yukawa couplings; their main effects show up in
the electric dipole moments (EDMs) of the neutron and of the charged leptons.
For $A_0 \sim m$ and $M_{1/2} = 400$ GeV, the contribution of
$|\mbox{Im}\, (\delta^d_{LR})^{11}|$ to the neutron EDM is under
control, while the contribution of $|\mbox{Im}\, (\delta^u_{LR})^{11}|$
provides a strong constraint (this is however mainly due to the fact that
the charge assignment predicts $A^{11}_u / A_0 \sim \lambda^6 \gg m_u / m_t$).
Still this constraint becomes weaker at high
$m$ as well as for large values of $M_{1/2}$ at low $m$.

We conclude that there are two regions in the (GUT-scale) MSSM parameter
space where the constraints associated with flavour and CP violating
processes are significantly weakened, both in the quark and lepton sectors:
(i) a low sfermion mass / high gaugino mass region, with
$m \lesssim 500$ GeV and $M_{1/2} \gtrsim 800$ GeV; and
(ii) a high sfermion mass region, with $m \gtrsim 10$ TeV and
practically no constraint on $M_{1/2}$. It is essentially flavour violation
in the lepton sector that requires high values of $M_{1/2}$ at low $m$.
Unless $m$ or $M_{1/2}$ are pushed towards very large values, however,
a strong suppression of off-diagional entries in the sfermion mass matrices
is still needed, especially in the $1$-$2$ sector. Splittings among the diagonal
entries are also constrained, but the allowed level of non-degeneracy
strongly depends on the Yukawa structure.


\section{Inverted sfermion mass hierarchy and $D$-term supersymmetry breaking}

However, if the supersymmetric FCNC and CP problems appear
to be less stringent in the previous regions, they  are less appealing
from the point of view of naturalness -- not even mentioning the possibility
of detecting superpartners at the LHC. As is well-known indeed,
in the MSSM the weak scale is determined by the following relation:
\beq
  \frac{1}{2}\, M_Z\ =\ - \mu^2 + \frac{m^2_{H_d} - m^2_{H_u} \tan^2 \beta}{\tan^2 \beta}\
  \simeq\ - \mu^2 - m^2_{H_u}\ ,
\label{eq:Mz}  
\eeq
where the last equality holds for $\tan \beta \gtrsim 10$. Naturalness requires
the absence of fine-tuned cancellations in the right-hand side of~\cite{ELENNAZW}
Eq.~(\ref{eq:Mz}).
Since $m^2_{H_u}$ receives large corrections from third generation sfermions
and gluinos, these should be light enough to keep  fine-tuning in the Higgs potential
still ``reasonable'' (say at the level of $1 \%$). The naturalness criterion then
appears to select the region of the parameter space where the FCNC problem
is maximized.

There is, however, a potential compromise, namely the so-called ``inverted
hierarchy'' scenario~\cite{DUPOSA,COKANE}, with heavy first two generation
squarks and sleptons in order to suppress FCNC and CP violating processes,
and light third generation sfermions and gluinos, in order to maintain
the fine-tuning at an acceptable level. In this section, we study a specific
realization of this scheme in the framework of gauge anomalous
horizontal $U(1)$ models~\cite{anomalous}.

\begin{figure}
\begin{center}
  \psfig{figure=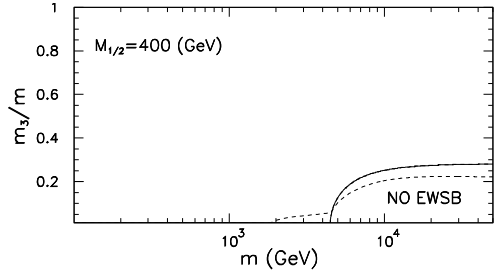,width=0.48\textwidth}
\end{center}
\caption{Exclusion curves in the ($m$, $m_3/m$) plane (assuming
$m_1 = m_2 \equiv m$) corresponding to proper
radiative electroweak symmetry breaking (solid line) and to the requirement
that the physical third generation sfermion masses, as well as the Higgs mass,
do not fall below the experimental limit (dashed line). \hfill
\label{fig:ftgg}}
\end{figure}

Before doing so, let us review the constraints that apply to the
inverted hierarchy scenario. As first pointed out in Ref.~\cite{ARMU},
large two-loop renormalization group effects associated with the
heavy first scalar generations could drive the squared masses
of the third generation squarks and sleptons negative.
Schematically, one has:
\beq
  16 \pi^2\, \frac{d\, m^2_3}{d \ln \mu}\ =\ 6\, y^2_3 m^2_3\, -\, \sum_i c_i\, g^2_i M^2_i\,
    +\, \frac{1}{16 \pi^2} \sum_i c'_i\, g^4_i m^2\ +\, \cdots\ .
\eeq
This contribution has to be compensated for by the contribution
of gauginos and by the initial value of the third
generation mass $m_3$. Avoiding tachyonic scalars then
implies a lower bound on the GUT-scale ratio $m_3 / m$
(where $m_1 \approx m_2 \approx m$), as a function of 
$M_{1/2}$ and $m$. An even stronger bound is obtained,
for large $m$ values, by requiring that the lightest Higgs mass
be larger than its experimental limit. We point out that actually
a more stringent constraint comes from the requirement of correct
electroweak symmetry breaking.
This is shown in Fig.~\ref{fig:ftgg}, in which one can see that, for
$m \gtrsim 5$ TeV, the ratio $m_3/m$ is constrained to be larger than
$0.2-0.3$ for radiative electroweak symmetry breaking
to be possible. A smaller value of $M_{1/2}$ (which has been set
to $400$ GeV in Fig.~\ref{fig:ftgg}) would not change this lower bound,
but it would also apply to lower values of $m$.

Let us now consider an explicit realization~\cite{NEWR}
of the inverted hierarchy scenario based on
$D$-term supersymmetry breaking~\cite{BIDU} in the framework
of gauge anomalous horizontal $U(1)$ models~\footnote{The fine-tuning issue
in such a scenario has been discussed in Ref.~\cite{BRASA}.}. Neglecting corrections
associated with non-canonical kinetic terms from the K\"ahler potential,
we can write e.g. the squark doublet soft supersymmetry breaking masses as:
\beq
  (m^2_Q)^{AB}\ =\ q_A\, \delta^{AB} m^2_D\ +\ C^{AB}_Q
  \lambda^{|q_A - q_B|}\, m^2_F\ ,
\label{eq:m2_Q}
\eeq
and similarly for the other squark and slepton soft masses. The first term
in Eq.~(\ref{eq:m2_Q}) is the contribution of the anomalous $D$-term
(with $m^2_D \equiv g <D>$), and the second term the contribution
of the $F$-terms. We assume a supergravity scenario leading to
a hierarchy of the contributions $m^2_F \ll m^2_D$, with however
$m_F / m_D \geq 0.3$ in order to satisfy the constraints discussed above.
The inverted hierarchy
scenario is then straightforwardly implemented by assigning zero
horizontal charges to the third generation superfields whose scalar
components couple most strongly to the Higgs bosons, namely
$\tilde t_L$, $\tilde t_R$ and $\tilde b_L$,
and strictly positive horizontal charges to the other quark
and lepton superfields~\cite{NEWR}. This condition is fulfilled
by the charge assignment used in section~\ref{sec:FCNC} to
describe the quark and lepton masses and mixings.

\begin{figure}
\begin{center}
  \psfig{figure=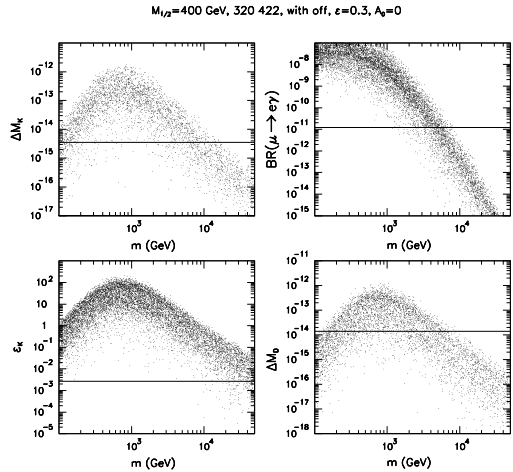,width=0.48\textwidth}
\end{center}
\vskip .1cm
\caption{$\Delta m_K$ ({\it top left}), $\mbox{BR} (\mu \rightarrow e \gamma)$ ({\it top right}),
$\epsilon_K$ ({\it bottom left}) and $\Delta m_D$ ({\it bottom right}) as a function
of $m_D$, in the $D$-term supersymmetry breaking scenario with
$m_F / m_D = 0.3$, $A_0 = 0$ and $M_{1/2} = 400$ GeV. 
Order one coefficients in the Yukawa matrices and in the off-diagonal entries
of the sfermion mass matrices are varied randomly in the range $[0.3,3]$. \hfill
\label{fig:320422r3g4}}
\end{figure}

In view of the FCNC problem, the main virtue of the
$D$- term-induced inverted hierarchy scenario is to suppress the
off-diagonal entries in sfermion mass matrices, since a
suppression factor of $m^2_F / m^2_D$ comes in addition to the
power of $\lambda$ associated with the breaking of the horizontal symmetry
(this however assumes a canonical K\"ahler potential).
The diagonal entries, however, are strongly split. Fig.~\ref{fig:320422r3g4}
shows $\Delta m_K$, $\epsilon_K$, $\mbox{BR} (\mu \rightarrow e \gamma)$
and $\Delta m_D$ as a function of $m_D$ for model 1 of Ref.~\cite{CHKOLAPO},
assuming $m_F / m_D = 0.3$, $A_0 = 0$ and $M_{1/2} = 400$ GeV.
In addition to the Yukawa couplings, the coefficients of the off-diagonal
entries $C^{AB}_{Q,U,D,L,E}$ in the sfermion mass matrices are varied
randomly in the range $[0.3,3]$.
Supersymmetric contributions to $\Delta m_K$,
$\mbox{BR} (\mu \rightarrow e \gamma)$ and $\Delta m_D$ are under
control in the region $m_D \gtrsim 5$ TeV; however, $\epsilon_K$ still requires
values of $m_D$ in the several tens of TeV range, corresponding to even
larger values of the sfermion masses (one has e.g.
$m_{Q_1} = m_{U_1} = 1.7 m_D$ and $m_{D_1} = 2 m_D$).
This is due to the fact that phases in the Yukawa
couplings and in off-diagonal entries are taken to be random in our analysis,
and can therefore be large. Also, due to the 2-loop renormalization group
effect mentioned above, the stops, left sbottom and right stau are
light and can give significant contributions to $1-2$ flavour changing
processes due to the flavour mixing present in the fermion and
sfermion mass matrices. This also implies that some scalar superpartners
could be accessible at the LHC, even though the first two generations
are very heavy.

%


\section{Discussion and conclusions}

We have studied the contraints on non-flavour-blind soft supersymmetry
breaking terms coming from FCNC and CP violating processes
in the presence of hierarchical Yukawa couplings. 
These constraints are significantly weakened in the ``low'' sfermion mass
/ high gaugino mass region, with $m \lesssim 500$ GeV and
$M_{1/2} \gtrsim 800$ GeV (where both $m$ and $M_{1/2}$ are
GUT-scale parameters), as well as in
the high sfermion mass region, with $m \gtrsim 10$ TeV and
essentially no restriction on $M_{1/2}$.
Unless $m$ or $M_{1/2}$ are pushed towards very large values, however,
a strong suppression of off-diagonal entries in the sfermion mass matrices
is still needed, especially in the $1$-$2$ sector. Splittings among the diagonal
entries are also constrained, but the allowed level of non-degeneracy
strongly depends on the Yukawa structure.
Quark textures for which the CKM matrix comes mainly from
the up quark sector (see e.g. Ref.~\cite{NIRA}), while the mixing
in the left-handed down quark sector is strongly suppressed,
would allow for larger splittings than the ``generic'' hierarchical textures
that we have considered. A similar statement can be made about the
charged lepton Yukawa texture.

We have also studied the inverted sfermion mass hierarchy scenario
in the framework of gauge anomalous horizontal $U(1)$ models
with $D$-term supersymmetry breaking. In such a scheme,
the diagonal entries of the sfermion mass matrices are strongly split,
but the off-diagonal entries are suppressed. FCNC and CP
constraints can be satisfied at the price of very heavy first generation
squarks, with masses in the few $10$ TeV range. However, the situation
could improve with different quark Yukawa textures arising e.g. from
less minimalistic anomalous $U(1)$ models.


\section*{Acknowledgments}
It is pleasure to thank the organizers of the XLth Rencontres de Moriond for 
creating a pleasant and stimulating atmosphere at the conference.
This work has been supported in part by the RTN European Program 
MRTN-CT-2004-503369. P.H.Ch. and S.P. were supported by the Polish State 
Committee for Scientific Research Grant 2 P03B 129 24 for 2003-2005.



\section*{References}
\begin{thebibliography}{99}
%
\bibitem{ELNA} J.R. Ellis and D.V. Nanopoulos, \Journal{\PLB}{110}{44}{1982}.
%
\bibitem{NISE} Y. Nir and N. Seiberg,
\Journal{\PLB}{309}{337}{1993}.
%
\bibitem{DUPOSA} E. Dudas, S. Pokorski and C.A. Savoy,
\Journal{\PLB}{369}{255}{1996}.
%
\bibitem{COKANE} A.G. Cohen, D.B. Kaplan and A.E. Nelson,
\Journal{\PLB}{388}{588}{1996}.
%
\bibitem{CHKOLAPO} P.H. Chankowski, K. Kowalska, S. Lavignac and S. Pokorski,
\Journal{\PRD}{71}{055004}{2005}.
%
\bibitem{FRONI} C.D. Frogatt and H.B. Nielsen, \Journal{\NPB}{147}{277}{1979}.
%
\bibitem{ALFEMA} G. Altarelli, F. Feruglio and I. Masina,
\Journal{\JHEP}{0301}{035}{2003}.
%
\bibitem{HAKORA} L.J. Hall, V.A. Kostelecky and S. Raby,
\Journal{\NPB}{267}{415}{1986}.
%
\bibitem{BOMA} F. Borzumati and A. Masiero, \Journal{\PRL}{57}{961}{1986}.
%
\bibitem{BRIBMU} A. Brignole, L.E. Ibanez and C. Munoz, \Journal{\NPB}{422}{125}{1994};
D. Choudhury, F. Eberlein, A. Konig, J. Louis and S. Pokorski,
\Journal{\PLB}{342}{180}{1995};
P. Brax and C.A. Savoy, \Journal{\NPB}{447}{227}{1995}.
%
\bibitem{CHLEPO} P. H. Chankowski, O. Lebedev and S. Pokorski, 
\Journal{\NPB}{717}{190}{2005}.
%
\bibitem{GAGAMASI} F. Gabbiani, E. Gabrielli, A. Masiero and L. Silvestrini,
\Journal{\NPB}{477}{321}{1996}.
%
\bibitem{ELENNAZW} J.R. Ellis, K. Enqvist, D. V. Nanopoulos and F. Zwirner,
\Journal{\MPLA}{1}{57}{1986};
R. Barbieri and G.F. Giudice, \Journal{\NPB}{306}{63}{1988}.
%
\bibitem{anomalous}
L.E. Ibanez and G.G. Ross, \Journal{\PLB}{332}{100}{1994};
P. Binetruy and P. Ramond, \Journal{\PLB}{350}{49}{1995};
V. Jain and R. Shrock, \Journal{\PLB}{352}{83}{1995};
E. Dudas, S. Pokorski and C.A. Savoy, \Journal{\PLB}{356}{45}{1995};
Y. Nir, \Journal{\PLB}{354}{107}{1995};
P. Binetruy, S. Lavignac and P. Ramond, \Journal{\NPB}{477}{353}{1996}.
%
\bibitem{ARMU} N.~Arkani-Hamed and H.~Murayama, \Journal{\PRD}{56}{6733}{1997}.
%
\bibitem{NEWR} A.E. Nelson and D. Wright, \Journal{\PRD}{56}{1598}{1997}.
%
\bibitem{BIDU} P. Binetruy and E. Dudas, \Journal{\PLB}{389}{503}{1996};
G.R. Dvali and A. Pomarol,  \Journal{\PRL}{77}{3728}{1996}.
%
\bibitem{BRASA} P. Brax and C.A. Savoy, \Journal{\JHEP}{0007}{048}{2000}.
%
\bibitem{NIRA} Y. Nir and G. Raz, \Journal{\PRD}{66}{035007}{2002}.
%
\end{thebibliography}
\end{document}